# TITLE

# Ag colloids and arrays for plasmonic non-radiative energy transfer from quantum dots to a quantum well

AUTHOR NAMES

*Graham P. Murphy‡[1*], John J. Gough‡[1], Luke J. Higgins[1], Vasilios D. Karanikolas[1], Keith M. Wilson[1], Jorge A. Garcia Coindreau[1], Vitaly Z. Zubialevich[2], Peter J. Parbrook[2], and A. Louise Bradley[1]*

AUTHOR ADDRESS

[1] School of Physics and CRANN, Trinity College Dublin, College Green, Dublin 2, Ireland

[2] Tyndall National Institute and School of Engineering, University College Cork, Lee Maltings, Prospect Row, Cork, Ireland

ABSTRACT

Ag nanoparticles in the form of colloids and ordered arrays are used to demonstrate plasmon-mediated non-radiative energy transfer from quantum dots to quantum wells with varying top barrier thicknesses. Plasmon-mediated energy transfer efficiencies of up to ~25% are observed with the Ag colloids. The distance dependence of the plasmon-mediated energy transfer is found to follow the same $d^{-4}$ dependence as the direct quantum dot to quantum well energy transfer. There is also evidence for an increase in the characteristic distance of the interaction, thus indicating that it follows a Förster-like model with the Ag nanoparticle-quantum dot acting as an enhanced donor dipole. Ordered Ag nanoparticle arrays display plasmon-mediated energy transfer efficiencies up to ~21%. To explore the tunability of the array system, two arrays with different geometries are presented. It is demonstrated that changing the geometry of the array allows a transition from overall quenching of the acceptor quantum well emission to enhancement, as well as control of the competition between the quantum dot donor quenching and energy transfer rates.

**Introduction**

Hybrid semiconductor structures of quantum dots (QDs) and quantum wells (QWs) can play a large role in optoelectronic devices for photovoltaics[1–3], white light generation[4–6], and color conversion[7–10]. QDs provide the capability to collect sunlight over the full visible spectrum, and into the infrared[11–13], and InGaN/GaN QWs are also studied as a promising avenue for solar cell technology[14,15]. Utilizing QDs for solar cells is a large and diverse research field[16]. This work will focus on the area of non-radiative energy transfer (NRET) of energy absorbed by QDs[17–19]. Förster-type NRET is a dipole-dipole interaction between resonant donor and acceptor species and has a well-defined distance dependence[20]. NRET has been demonstrated in a large range of material systems, with potential for LED[9,21–25] and light harvesting applications[2,3,13,26–33]. QDs possess excellent optical properties such as broadband absorption with narrow and tunable emission profiles, giving them a distinct advantage over organic dyes[34,35]. However, typically they suffer from poor electrical properties, requiring, for example, additional charge transport layers[36]. Coupling to a system such as a QW would be highly advantageous in terms of carrier transport. Radiative transfer of energy absorbed by the QDs to a QW is inherently lossy. Single-step energy transfer (ET), such as that provided by NRET, is a preferred route for transferring the energy as it has the potential to be more efficient. Studies of NRET from QDs to other various semiconductor materials have been reported. Lu *et al.*[1] demonstrated NRET from PbS QDs to InGaAs QWs with an efficiency of 30%. They proposed a solar cell design utilizing NRET and postulated an efficiency of approximately 19% assuming unity quantum yield for the QDs. Further work by that group[30] demonstrated NRET from PbS QDs to Si nanowires with efficiencies in the range of 15-38%, and showed that the photocurrent due to NRET can be 3 times that due to direct light absorption in the Si nanowires for excitation near the QD absorption peak. Chanyawadee *et al.*[2,13] micro-patterned hybrid QD-GaAs based p-i-n devices to bring QDs into close proximity with the intrinsic region of p-i-n heterostructures, and thereby benefit from efficient NRET from the QDs. A six-fold increase of the photocurrent conversion efficiency for a patterned device with CdSe/CdS QDs and a GaAs multi-QW stack[2], and a three-fold increase for CdTe QDs and bulk GaAs device[13] were observed. Nizamoglu *et al.*[3] found NRET efficiencies of nearly 70% from CdSe/ZnS QDs to InGaN QWs with a center-to-center separation of ~3.8 nm, with the small separation required due to the strong distance dependence of NRET. As can be seen in a number of the examples cited, it is possible to combat the strong distance dependence with nanopillar architectures[31], graded nanocrystal layers[27,32,33], or buried QD nanostructures[1], and micro-patterning of a p-i-n device[2,13] which try to maximize the number of donors in close proximity to the acceptor by increasing the available acceptor surface area. These methods introduce additional design complexity and, in some cases, fabrication related defects. Another method to circumvent the limited length scale takes advantage of localized plasmon resonances (LSPs) of metal nanoparticles (MNPs). Plasmon-mediated NRET has been used to enhance the non-radiative interaction; enhancement of the NRET rate, efficiency, and range has been established for different geometries with a variety of emitters and metallic structures[37–47]. Within this group, enhancement of the QD emission from a thick (~ 80 nm) layer of QDs via NRET from QWs utilizing arrays of Ag nanoparticles (AgNPs) has been observed[48]. The potential to increase the interaction between the QW and QDs dispersed in a thicker layer is promising for device applications.

Plasmonic MNPs have been extensively considered as a route towards enhancement of photovoltaic devices[49,50]. MNPs can be used as sub-wavelength scattering elements to trap propagating waves into a thin semiconductor layer, and are also useful as sub-wavelength antennas where the plasmonic near-field can be coupled to the semiconductor, thereby increasing its effective absorption cross section. AgNPs have been shown to improve the efficiency of wafer-based Si solar cells, particularly in the near infrared where Si is a poor absorber[51]. Lee *et al*.[52] demonstrated an overall increase of the photocurrent by a factor of seven, most obviously observed at the plasmon resonance of the annealed Au films on $TiO_2$. The photo-response following the plasmon resonance is widely observed for Ag and Au[53–55]. Enhanced absorption and emission of QDs using plasmonics has also been demonstrated with a view towards photovoltaics[56,57]. Therefore, in a light harvesting system the plasmonic element could potentially not only be used as a solar concentrator but could perform a second important role in enhancing ET between the QD layer and the QW.

Plasmon-mediated ET from a high energy QD (2.75 eV / 450 nm) into a QW has yet to be demonstrated. Herein we provide evidence of plasmon-mediated ET from QDs to QWs using two different plasmonic systems, small colloidal AgNPs with a diameter of 6 nm and ordered arrays of AgNPs with nanostructure dimensions of ~ 90 nm and height of 40 nm. In the first system studied, QDs and colloidal AgNPs have been spin coated on three QWs with different top barrier thicknesses. The varying top barrier thickness allows us to probe the distance dependence of the direct and plasmon-mediated ET. The direct QD to QW ET, in the absence of any metal NPs, is seen to have the characteristic $d^{-4}$ Förster-like distance dependence to a 2D acceptor. This is a clear demonstration that it is the dimensionality of the acceptor that drives the distance dependence, which has previously been theoretically predicted[58]. Following on from this, the demonstration of surface plasmon-mediated ET from QDs to QWs is presented for AgNPs mixed with the QDs before being spin cast onto the QW. The distance dependence of the surface plasmon-mediated ET to QWs with different barrier thicknesses is also found to follow a $d^{-4}$ distance dependence, but with evidence of an increased energy transfer range, to the authors knowledge we believe this to be the first demonstration of surface plasmon-mediated ET from a QD into a QW. In the second system studied, arrays of Ag were fabricated onto a 3 nm barrier QW sample and again surface plasmon-mediated ET was observed, with a doubling of the plasmon-mediated ET rate. The AgNP arrays are shown to be highly tunable, with a large difference in the QD and QW quenching and ET rates observed for the two arrays studied. Consequently, it is demonstrated that by changing the geometry of the array, through the nanostructure shape and pitch, it is possible to go from a situation of overall quenching of the acceptor QW emission to one of emission enhancement relative to the QW alone, which could be highly advantageous for applications.

**Methods**

Colloidal CdSeS/ZnS QDs and AgNPs were purchased commercially from Sigma Aldrich and PlasmaChem, respectively. The QDs were used as purchased as a stock solution. The AgNPs were provided in powdered form and dissolved in toluene at a concentration of 10 mg/4mL. This was then used as a stock solution. The dispersions in PMMA were made by mixing 20 μL of stock QDs, 40 μL stock AgNPs, and 40 μL of 0.1 % w/w PMMA and sonicating for 20 seconds to ensure they were evenly dispersed. To produce the reference layers either the QDs or AgNPs were removed and replaced with the same volume of toluene to preserve the same PMMA:sample ratio. For example to produce the QD reference layer 20 μL of QDs, 40 μL of toluene, and 40 μL of PMMA were dispersed together. For the ordered array samples 5 μL of stock QDs and 95 µL of PMMA were dispersed together. The samples were spin cast onto the substrates at 5000 RPM for 60 seconds, producing layers approximately 6 nm thick.

The arrays are fabricated using electron beam lithography. A focused beam of electrons is used to draw a pattern of boxes or discs in an electron sensitive resist layer. The resist layer in this case is a 100 nm thick layer of PMMA. Methyl-isobutylketone (MIBK) is used to remove the exposed regions within the resist. After this development process metal is then deposited everywhere by metal evaporation. First 5 nm of Ti is deposited as an adhesion layer, followed by 35 nm of Ag. The remaining resist is removed by a lift off process in acetone, taking the metal from unexposed regions off with it. Therefore only metal from areas that were exposed by the electron beam remain.

A calculation of the concentration of the QDs can be done based on the number of QDs per mg quoted by the manufacturers; this is $1.2 \times 10^{15}$ QDs per mg. The stock solution is 5 mL at 1 mg/mL; therefore the total number of dots in the stock solution is $6 \times 10^{15}$. Aliquots of 20 µL are used for mixing with the PMMA and AgNPs, in each 20 µL of the stock solution there must be $2.4 \times 10^{13}$ QDs. This aliquot is made up to 100 µL when mixing with the PMMA and AgNPs giving a concentration of $2.4 \times 10^{13}$ per 100 µL or $2.4 \times 10^{17}$ $L^{-1}$. Converting to cubic meters gives a concentration of $2.4 \times 10^{20}$ $m^{-3}$. For the Ag array samples with the reduced concentration only 5 µL of the QD solution is used in 100 µL, following the same calculation gives a concentration of $6 \times 10^{19}$ $m^{-3}$.

Absorption and extinction spectra were measured using a Cary 50 UV-vis spectrophotometer. PL spectra were obtained from an Andor Shamrock sr-303i spectrometer with an Andor Newton 970EMCCD. This spectrometer is fiber coupled to an output port of a PicoQuant Microtime 200 Fluorescence Lifetime Imaging Microscope (FLIM) which is used to measure time resolved photoluminescence (TRPL). The samples were excited through a 40x objective using picosecond laser pulses at 405 nm with a repetition rate of 10 MHz, emission was collected back through the same objective. The PL decays were recorded over an 80 x 80 $\mu m^2$ for the samples involving the colloidal AgNPs, smaller 20 x 20 $\mu m^2$ areas were used for the Ag arrays. Spectral filtering is achieved using a combination of narrow-band and broadband emission filters. The QD emission is selected using a 450 nm filter with full-width-half-maximum of (8 ± 2) nm, the QWs are selected with a 550 nm filter with full-width-half-maximum of (70 ± 5) nm. PL decays were fitted with a two-exponential decay function. The average lifetime, $\tau_{av}$, is calculated from the intensity-weighted mean: $\tau_{av} = (A_1\tau_1^2 + A_2\tau_2^2)/(A_1\tau_1 + A_2\tau_2)$. The average lifetime quoted for the Ag colloid system is the average of 5 measurements on each sample. For the array samples one lifetime measurement is taken on the arrays and one off, due to the highly ordered nature of the arrays it is possible to return to the same area for subsequent measurements. The average decay rate is then easily determined as the inverse of this average lifetime. All samples were fabricated and measured at room temperature.

**Results and Discussion**

CdSeS/ZnS QDs with a diameter of (6.0 ± 0.5) nm and colloidal AgNPs also with a diameter of (6.0 ± 0.5) nm (both in toluene) are dispersed separately and together in a 0.1% w/w solution of PMMA in toluene and spin cast onto metalorganic vapour phase epitaxy (MOVPE) prepared QWs with differing GaN barrier thickness. This produces layers of QDs alone and a mixed layer of AgNPs and QDs with a thickness of approximately 6 nm, confirmed by atomic force microscopy (AFM) (Supporting Information figures S1(a) and 1(b). Reference layers are spin cast onto bulk GaN. Figure 1(a) shows the extinction spectrum of the AgNPs in a PMMA solution and in a layer spin cast onto a quartz substrate. Clearly in the layer the overall extinction is reduced, and the LSP resonance appears to be broadened. This is indicative of aggregation of the AgNPs. Evidence of aggregation is also seen in the scanning electron microscope (SEM) images of a layer of the AgNPs spun onto a silicon wafer (to prevent charging during SEM imaging). Clearly visible in the inset of figure 1(a) are large aggregations of AgNPs, up to tens of microns in size.

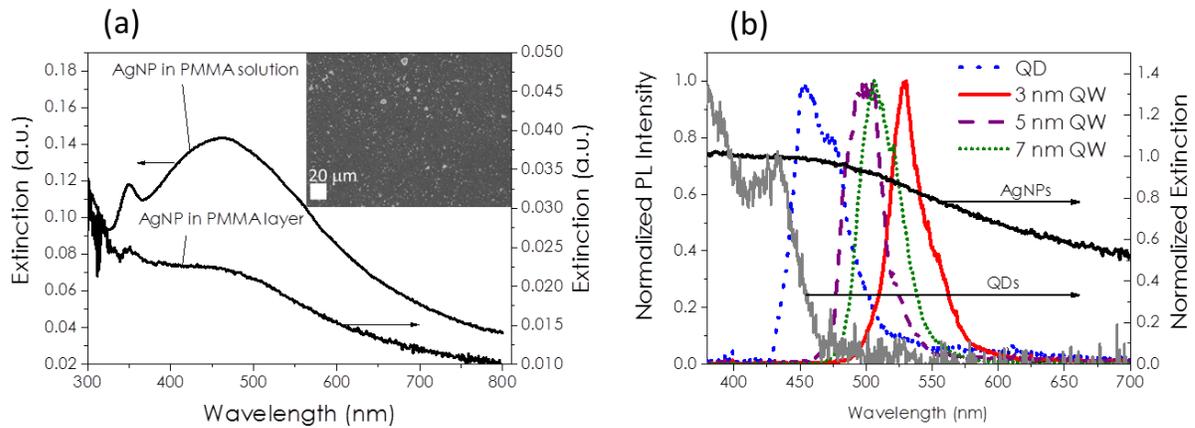

**Figure 1.** (a) Extinction spectra of the AgNPs dispersed in a PMMA solution and in a PMMA layer spun onto quartz. Inset: SEM image of AgNPs in a PMMA layer spun onto a silicon substrate. This shows approximately 10 % coverage with the AgNPs. (b) PL spectra of the QDs dispersed in PMMA and the 3, 5, and 7 nm barrier QW (left axis). The normalized absorption spectrum of the QDs (grey) and the extinction spectrum of the AgNPs (black) are also shown (right axis).

The photoluminescence (PL) spectra of the QDs dispersed in PMMA and three QWs are shown in figure 1(b). The QD concentration is $2.4 \times 10^{20}$ m$^{-3}$ and the PL spectrum is centered on 450 nm. The QWs have different center peak wavelengths for the three top barrier thicknesses, 530 nm for the 3 nm barrier, 500 nm for the 5 nm barrier, and 510 nm for the 7 nm barrier. There is no significant overlap of the QW emission with the QD absorption. Also shown is the AgNP extinction spectrum to indicate the overlap of the PL spectra with the LSP peak. Single exciton generation is ensured by pumping the QDs with a relatively low excitation density, less than 1 µJ/cm$^2$ (see supporting information figure S2).

*Energy transfer using Ag nanoparticles*

The main signatures of non-radiative ET in a system are decreases in the lifetime and the PL emission of the donor species. The lifetime of the donor is therefore measured in a QD layer on a GaN substrate (QD only), in the presence of the QW (QD-QW), in the presence of AgNPs on a GaN substrate (QD-Ag), and finally, in the presence of both the AgNPs and the QW (QD-Ag-QW). PL decays of these four samples are shown in figure 2(a), 2(b) and 2(c) for the 3 nm, 5 nm and 7 nm barrier QW, respectively. For each QW barrier thickness there is a decrease in the QD lifetime in the presence of the AgNPs, with a further decrease of the lifetime in the presence of the QW, indicating plasmon-mediated ET. This is most obvious for the 3 nm barrier QW in figure 2(a). The changes in lifetime can be clearly seen in Figure 2(d)-(f) showing the average lifetime, of the QDs on each QW for the four samples. As mentioned above this average is taken from five scans across the sample, which helps to combat any issues relating to the AgNP aggregation. The small variance in lifetime (< 5%) measured across the sample shows that the NRET efficiency is not strongly dependent on where on the sample the emission is collected from. The quenching efficiency of the QDs in the presence of the AgNPs is calculated from

$$E_Q = 1 - \frac{\tau_{QD-Ag}}{\tau_{QD}} \qquad (1)$$

where $\tau_{QD}$ is the average lifetime of the QD-only sample and $\tau_{QD\text{-}Ag}$ is the average QD lifetime of the QD-Ag sample. The quenching efficiency is approximately 17 % for the QD-Ag sample.

Focusing on the decrease of the lifetime from QD only to the QD-QW structure, shown as the blue and green lines in Figure 2(a)-(c), it is evident that this interaction is strongest for the 3 nm barrier QW. This is expected as NRET is very strongly distance dependent. It has been theoretically shown that the dimensionality of the acceptor determines the distance dependence[58], therefore, since the QW is a 2D material a $d^{-4}$ dependence is expected.

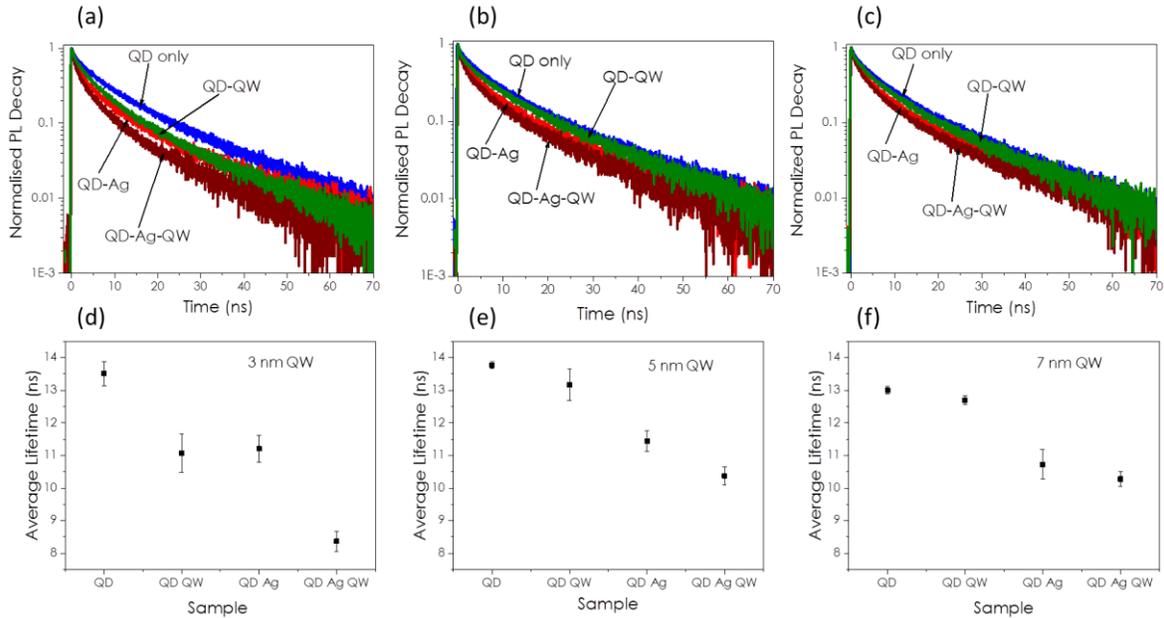

**Figure 2.** PL decays for the four samples, QD layer on GAN substrate (QD only, blue), the mixed QD and QNNP layer on a GaN substrate (QD-Ag, red), the QD layer on the QW (QD-QW, green), and the mixed QD and Ag NP layer on the QW (QD-Ag-QW, wine) for each QW top barrier thickness of 3 nm (a), 5 nm (b) and 7 nm (c). Average decay lifetimes for the four samples on the three different QWs 3 nm (d), 5 nm (e) and 7 nm (f).

The ET efficiency from QDs to QW in the absence of the AgNPs is given by

$$E_D = 1 - \frac{\tau_{QD-QW}}{\tau_{QD}} \qquad (2)$$

where $\tau_{QD}$ is defined as before and $\tau_{QD\text{-}QW}$ is the average QD lifetime of the QD-QW sample. The NRET efficiency when dealing with energy transfer to an acceptor plane has the form

$$E_{NRET} = \frac{1}{1 + kr^4} \tag{3}$$

where $k$ is a constant and $r$ is the center-to-center separation between the QD donor and the QW acceptor species. To verify that the experimentally observed dependence of the ET efficiency from QDs to QW is represented by $d^{-4}$, a logarithmic plot of the measured energy transfer efficiency versus distance is fitted to a power law (ax$^n$) in figure 3. A free fit (black line) yields an exponent n = -3.8 ± 0.9, close to the theoretical n = -4, indicating that the ET is Förster-like and governed by a $d^{-4}$ dependence. For a fit using (3) we obtain k = (3.0 ± 0.4) x $10^{-3}$ nm$^{-4}$.

In the presence of both AgNPs and the QWs, sample QD-Ag-QW, it is clear from figure 2 that there is further reduction in the lifetime of the QD. The surface plasmon-mediated ET efficiency is determined from

$$E_P = 1 - \frac{\tau_{QD-Ag-QW}}{\tau_{QD-Ag}} \tag{4}$$

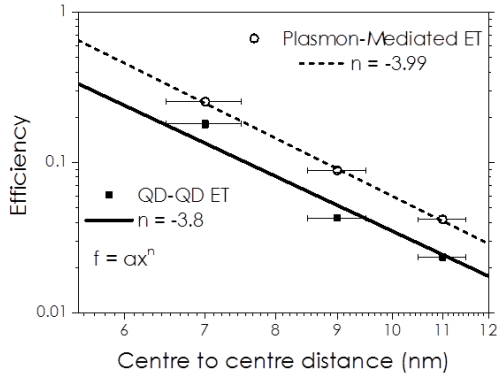

**Figure 3.** Logarithmic plot of the ET (solid squares) and plasmon-mediated ET (hollow circles) efficiencies. The black line (dash) is a free fit with a power law, producing n = -3.8 ± 0.9 and -3.9 ± 0.2 for the ET and plasmon-mediated ET respectively.

where $\tau_{QD\text{-}Ag}$ is the lifetime of the QDs in a mixed layer with the AgNPs on a GaN reference, and $\tau_{QD\text{-}Ag\text{-}QW}$ is the lifetime of the QDs in a mixed layer with the AgNPs on the QW. Figure 3 also shows the plasmon-mediated ET efficiency data represented on a logarithmic plot, it can be noted the efficiency in the presence of the AgNPs is larger at each distance than the direct ET. A free fit to a power law yields an exponent, n = -3.9 ± 0.2, indicating that surface plasmon-mediated ET follows the same $d^{-4}$ distance dependence as direct NRET. In this instance a fit with (3) yields $k$ = (1.6 ± 0.2) x $10^{-3}$ $nm^{-4}$. For a plane of dipole acceptors $k$ can be expressed as $k = 2/c_{Acc}\pi R_0^6$, where $c_{Acc}$ is the acceptor concentration and $R_0$ is the Förster radius, a characteristic distance at which the energy transfer efficiency is 50%[59–61]. A decrease in the value of $k$ can occur due to an increase in the acceptor concentration, or an increase in the Förster radius[61]. In this case the same QW is used for the QD-QW and the QD-Ag-QW samples, and therefore, the decrease in the value of $k$ can be attributed to an increase of the Förster radius, $R_0$. The $d^{-4}$ distance dependence coupled with increased characteristic energy ET distance (decrease in $k$) indicates that the combined QD-Ag can be considered as an enhanced donor dipole relative to the QD alone in a Förster-like model, similar to what has been reported for QD-QD plasmon-mediated NRET[46,62]. We believe this to be the first demonstration of this enhanced donor-dipole in systems where the ET is from a QD to a QW, and that this could have applications in enhancing light harvesting devices as mentioned in the introduction.

Ideally, an enhancement of the acceptor PL would be present in an ET system with two emitting species. However, no modification of the QW PL is observed in this study. This may be due to the close proximity of the mixed QD-AgNP layer to the acceptor QW. Increasing the barrier thickness would reduce quenching of the QW acceptor emission by the AgNPs[46]. An advantage of this colloidal system is that it utilizes simple processing techniques like spin coating that are easily scalable, however the placement of AgNPs is random which can influence the ET process and aggregation of the AgNPs needs to be overcome; having more isolated NPs could sharpen the LSP resonance giving an overall larger effect. In addition, having the AgNPs in an intermediate layer between the QDs and QW could provide even greater enhancement of the energy transfer efficiency[62,63].

*Energy transfer using Ag nanoparticle arrays*

The promising observation of plasmon-enhanced ET from the QDs to the QWs using the colloidal AgNPs motivated the consideration of an alternative approach using lithographically defined plasmonic arrays. Such arrays can provide better uniformity than the dispersed colloidal AgNPs for a range of nanostructures of differing dimensions and pitch. An array of nanodiscs (ND) and nanoboxes (NB) were selected, figures 4(a) and S3. Simulations of the extinction spectra were obtained using finite-difference time-domain (FDTD) simulation software using experimentally measured data[64], see supporting information for details. Both arrays show evidence for a peak in the extinction close to the QD emission wavelength, but the absorption and scattering are much larger for the ND array.

These NDs and NBs arrays were fabricated by electron beam lithography on the QW with the 3 nm top barrier thickness, as well as on GaN substrates for comparative reference measurements. SEM micrographs of the arrays on the QWs are shown in figure 4(b) and (c). The arrays consist of 200 x 200 units. The NDs (figure 4(b)) have a diameter of 90 nm with an 80 nm gap, giving a total area of 34 μm x 34 μm. The NBs (Figure 4(c)) consist of 100 nm x 100 nm boxes with a 200 nm gap giving a total area of 60 μm x 60 μm. The NDs and NBs had a total height of 40 nm consisting of 5 nm of Ti and 35 nm of Ag. As can be seen from the micrographs the individual structures are not perfectly defined due to the limitations of EBL, therefore they are just nominally NDs and NBs. However, the plasmonic response is also sensitive to the pitch, with the absorption and scattering decreasing with increasing gap between the individual nanostructure. Sample simulation data is shown in figure S3. Therefore, despite the poor definition of the individual nanostructures the interaction with the plasmonic array will be modified due to the change in pitch. To complete the structures the QDs in a PMMA solution, with a concentration of $6 \times 10^{19}$ m$^{-3}$, was subsequently spin cast onto the arrays.

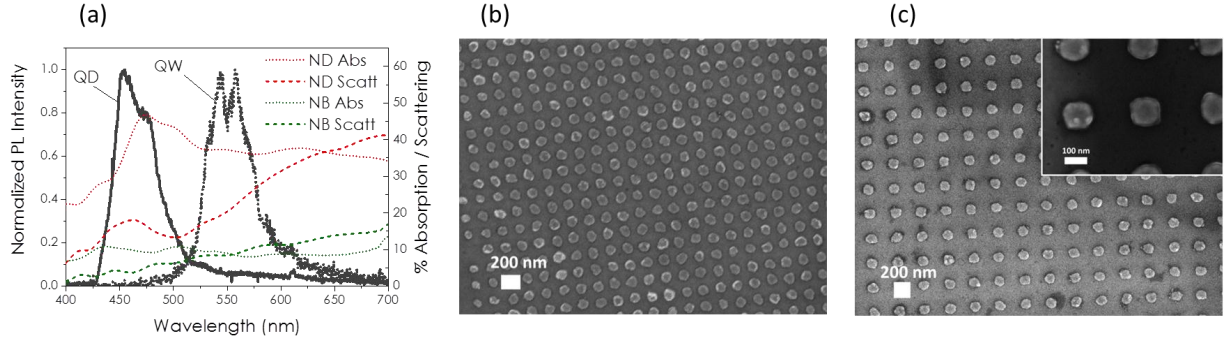

**Figure 4.** (a) Comparison between ND (red) and NB (green) simulation absorption (dot) and scattering (dash) spectra (see right axis). Normalized PL of the QD and QW (see left axis). (b) SEM micrographs of the NDs and (c) NBs, inset is a zoomed image.

The PL decays on and off the two arrays were measured over an area 20 x 20 μm$^2$ on the QW and reference GaN substrate, shown in figures 5(a) and 5(d). The average lifetimes are shown in figure 5(b) and 5(e). The QDs show a reduced lifetime on the QW compared with the GaN reference substrate indicating direct ET from the QDs to the QW. This direct QD to QW ET, $E_D$, is given by (2). The QDs show a shorter lifetime on the arrays on the QW than on GaN indicating the presence of plasmon-mediated ET. The quenching efficiency, $E_{Q,array}$, is given by (1) and the plasmon-mediated ET efficiency, $E_{P,array}$, is given by (4), where Ag now refers to the array and is replaced by ND or NB for the nominal nanodiscs or nanoboxes, respectively. All the extracted efficiencies and rates are presented in Table 1. While it can be noted that the plasmon-mediated ET efficiencies are not significantly different to the direct ET efficiency, the plasmon-mediated ET rate is approximately double that of the direct QD to QW energy transfer rate for both arrays. It can also be noted that the quenching efficiency, $E_{Q,NB}$, and the quenching rate of the QDs by the NB array, is much smaller than in the ND case. For the NB array the quenching and plasmon-mediated ET rates are comparable which enables them to compete and yield a slightly increased ET efficiency compared to the direct QD-QW case. For the ND array the quenching rate is dominant, resulting in a lower ET efficiency for the plasmon-mediated case compared to the direct NRET.

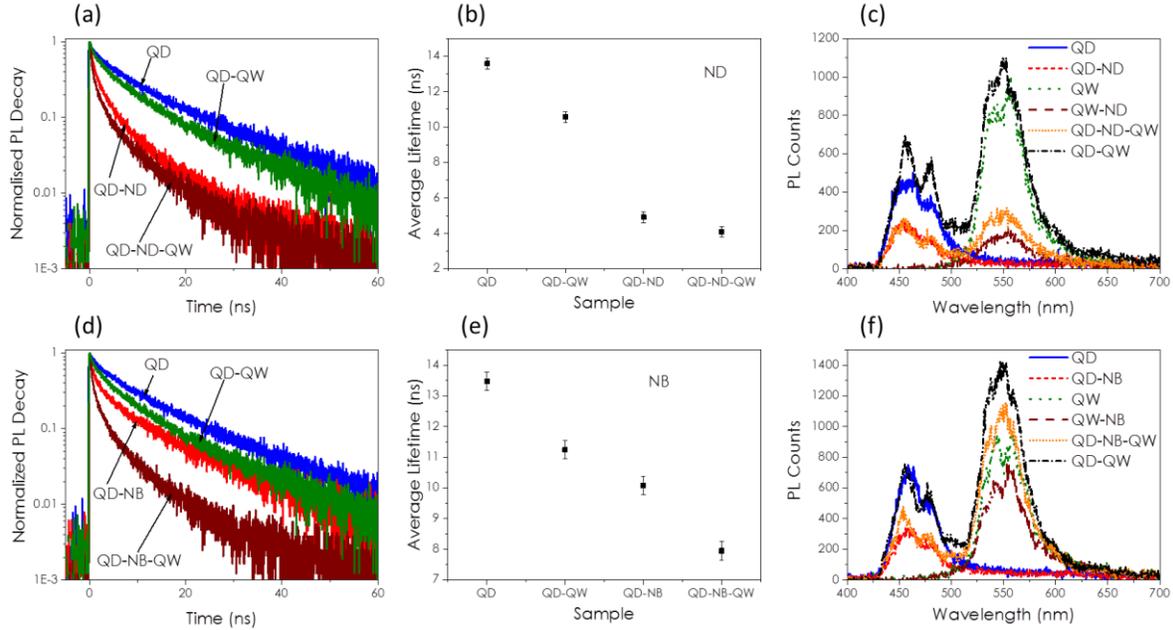

**Figure 5.** (a) [(d)] PL decays for four cases with the ND [NB] array. QD (blue), QD-ND (red), QD-QW (green) and QD-ND-QW (black). (b) [(e)] Average lifetimes from the PL decay of the QD on all four samples for the ND [NB] array. (c) [(f)] PL spectra for the ND [NB] array showing QD (blue), QD-ND (red), QW (green), QW-ND (wine), QD-ND-QW (orange) and QD-QW (black).

The PL spectra are shown in figure 5(c) and 5(f). The initial QD spectrum on the GaN substrate (blue) and QW emission (green) can be compared with the array structures. Taking firstly the ND array, there is clear emission quenching of the QDs due to the ND array on GaN (QD-ND), with very slight further quenching on the array decorated QW (QD-ND-QW). There is also very efficient quenching of the QW emission when it is decorated by the ND array (QW-ND, wine), but with the addition of the QDs the QW emission recovers (QD-ND-QW) above this quenched level. However, it does not recover beyond the initial QW emission.

This contrasts with the NB array for which the QW emission in the QD-NB-QW case is higher than the initial QW emission. It is the more efficient quenching in the case of the ND that leads to the greater reduction of the QW emission and prevents enhancement of the PL in the plasmon-mediated case. This can be attributed to the higher absorption component of the ND extinction, presented in figure 4(a). The QW emission shown in figure 5(f) is still not as large in the plasmon-mediated case as it is for the direct QD-QW NRET case. However, herein only two arrays have been considered, and the high level of tunability observed suggests that further improvement of the plasmon-mediated case could be achieved.

| Efficiency | QD-QW $E_D$ | QD-Array $E_{Q,Array}$ | Plasmon-Mediated $E_{P,Array}$ |
|---|---|---|---|
| ND | 0.22 ± 0.01 | 0.64 ± 0.04 | 0.17 ± 0.02 |
| NB | 0.17 ± 0.01 | 0.25 ± 0.01 | 0.21 ± 0.01 |
| **Rates** | **QD-QW** | **QD-Array** | **Plasmon-Mediated** |
| ND | (0.0210 ± 0.0007) ns$^{-1}$ | (0.130 ± 0.009) ns$^{-1}$ | (0.04 ± 0.01) ns$^{-1}$ |
| NB | (0.0148 ± 0.0005) ns$^{-1}$ | (0.0251 ± 0.0009) ns$^{-1}$ | (0.027 ± 0.001) ns$^{-1}$ |

**Table 1.** Summary of the efficiencies and rates for QD-QW ET, QD-array quenching, and plasmon-mediated ET.

The measurements suggest that adjustment of the structures to better control the trade-off between the quenching rates and the plasmon-mediated ET rate could allow for further tuning of the acceptor enhancement with arrays. Research on colloidal plasmon-mediated ET systems has shown that plasmon coupling to the donor has a larger influence on the ET process, allowing for the acceptor to be placed further from the plasmonic component, thereby reducing the direct quenching of the acceptor[46,62]. Therefore the QW barrier thickness is another parameter that can be tuned to move from a regime of quenching to one of enhancement[57,65]. A thicker QW barrier is also advantageous for device applications as it reduces non-radiative losses. Using a thicker layer of QDs would enable the QW to benefit from QDs at distances where direct non-radiative ET is negligible. Varying the spacing between the elements in the array and changing the size or shape of the individual units modifies the absorption and scattering profiles of the arrays, which can dramatically alter the emission of the QDs and QWs. As shown by way of example in figure S3 as the gap in the arrays is reduced the absorption and scattering increases. The dramatic changes in the strength of the absorption and scattering just by varying the gap size and leaving the dimensions of the arrays untouched show the scope for optimization in these array structures. A full optimization of the QW emission via plasmon-mediated ET would require a more complex model taking account of the distribution of the QDs within the layer deposited over the array and the distribution of dipole emitters within the QW under the array.

**Conclusions**

This work has demonstrated the distance dependence of QD to QW ET using CdSeS/ZnS QDs emitting at 450 nm and InGaN/GaN QWs with GaN barrier thicknesses of 3, 5, and 7 nm, respectively. The ET efficiency falls off with the characteristic $d^4$ distance dependence of Förster-like NRET to a plane, verifying theoretical work that showed that the dimensionality of the acceptor drives the distance dependence[58]. We have also reported on the demonstration of plasmon-mediated ET from QDs to a QW for the first time, using both colloidal AgNPs and Ag EBL fabricated arrays. With the AgNPs we investigated the distance dependence of the surface plasmon-mediated ET efficiency, and again a $d^{-4}$ distance dependence was determined, with evidence of an enhanced characteristic distance. No acceptor QW emission enhancement was observed in the colloidal system, however there is further room for optimization, both in the geometry of the system and in the quality of the AgNPs and their resulting LSP resonance. Within the Ag array system two different structures were investigated, denoted as NBs and NDs. The varying levels of acceptor enhancement for these two structures demonstrate the possibility to tune the QW emission. The ND structure proved to have the greatest interaction with the QDs and QW; however the resulting high level of quenching meant that the QW emission was only able to slightly recover in the plasmon-mediated case, leading to an overall reduction of the QW emission. For the NB array the quenching was more moderate, allowing for an enhancement of the QW emission in the plasmon-mediated ET case compared to the QW alone. Having demonstrated the potential for plasmon-mediated ET from QDs to QW, there are many parameters which can be adjusted to optimize the system. Such structures could be advantageous for light harvesting devices.

## ASSOCIATED CONTENT

**Supporting Information**.

This material is available free of charge via the Internet at http://pubs.acs.org."

The section includes AFM measurement of the PMMA layer thickness, power dependence of the lifetime and PL intensity for the QDs, and further details of the FDTD simulations.

## AUTHOR INFORMATION

**Corresponding Author**

*Address correspondence to gpmurphy@tcd.ie Tel: +353830067704

**Author Contributions**

The manuscript was written through contributions of all authors. All authors have given approval to the final version of the manuscript. ‡These authors contributed equally.

**Funding Sources**

This work was supported by Science Foundation Ireland (SFI) under grant number 10/IN.1/12975 and the National Access Programme Grant under grant number NAP 338, and enabled using facilities funded by Irish Higher Education Authority Programme for Research in Third Level Institutions Cycles 4 and 5 via the INSPIRE and TYFFANI projects. JJG and GPM acknowledge postdoctoral research fellowships from the Irish Research Council (RS/2011/287 and GOIPG/2013/680) and PJP a SFI Engineering Professorship (SFI/07/ EN/E001A).

ACKNOWLEDGMENT

We thank Robert O'Connell for the AFM measurements of the PMMA layers.

## SUPPORTING INFO

**AFM Measurement**

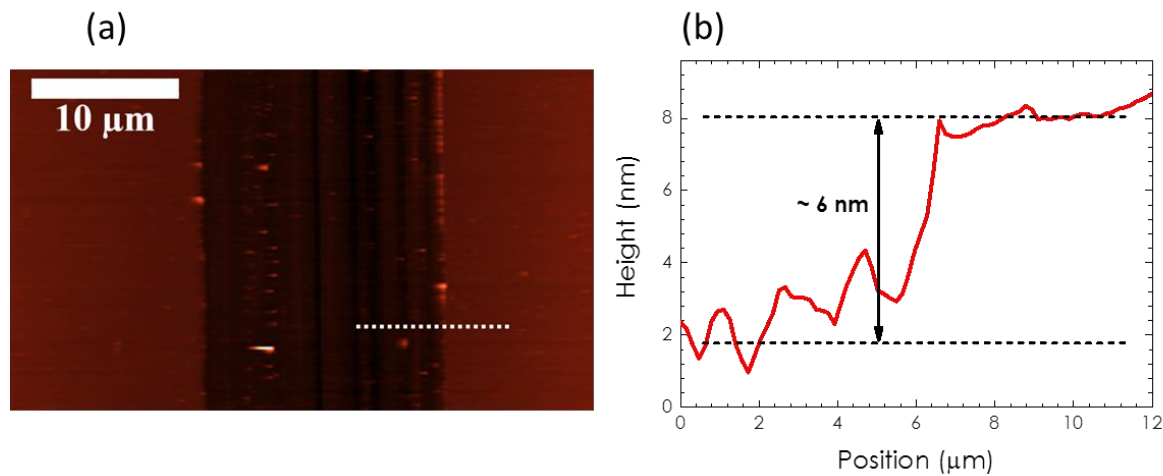

**Figure S1.** (a) AFM of a cut in the QD PMMA layer (b) Measurement of the step height of the layer.

**Power Dependence**

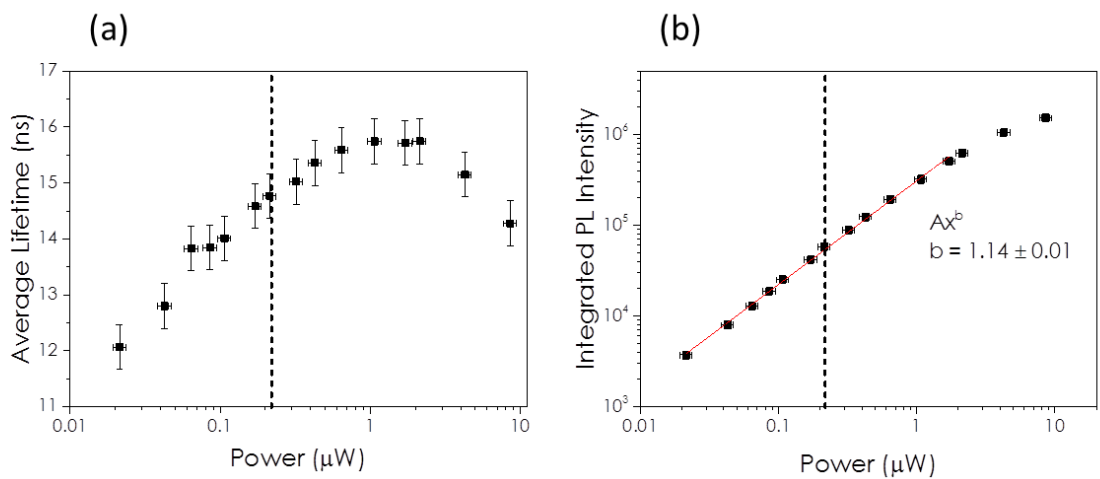

**Figure S2.** (a) Dependence of the QD lifetime on excitation power. Dashed line represents the power at which subsequent measurements were taken. (b) Dependence of the QD integrated PL intensity on excitation power. Solid red line is a fit to a power law with b = 1.14 ± 0.01, i.e. a linear increase with power, indicating only single exciton generation in each QD. The dashed line shows the power at which the measurements are taken.

The dependence of the QD average lifetime and PL intensity on excitation power is shown in Figure S2. The QD lifetime increases with increasing power (Figure S2a), and the PL intensity increases linearly with excitation power (Figure S2b) indicating single exciton generation in each QD. The fall-off in both the lifetime and PL intensity at high power is attributed to the onset of photobleaching. The dashed line shows the power at which subsequent measurements are taken. It was selected to ensure that the QDs are pumped in the linear regime, away from the onset of photobleaching effects

**FDTD Simulations**

The extinction (absorption + scattering) of the arrays were simulated using plane-wave excitation. The PMMA and GaN layers are modelled with a constant dielectric permittivity of $\varepsilon_{PMMA} = 2.2$ and $\varepsilon_{GaN} = 5.35$, respectively. The dielectric permittivity of the Ag and Ti materials are included using experimentally measured data from Palik.[64]

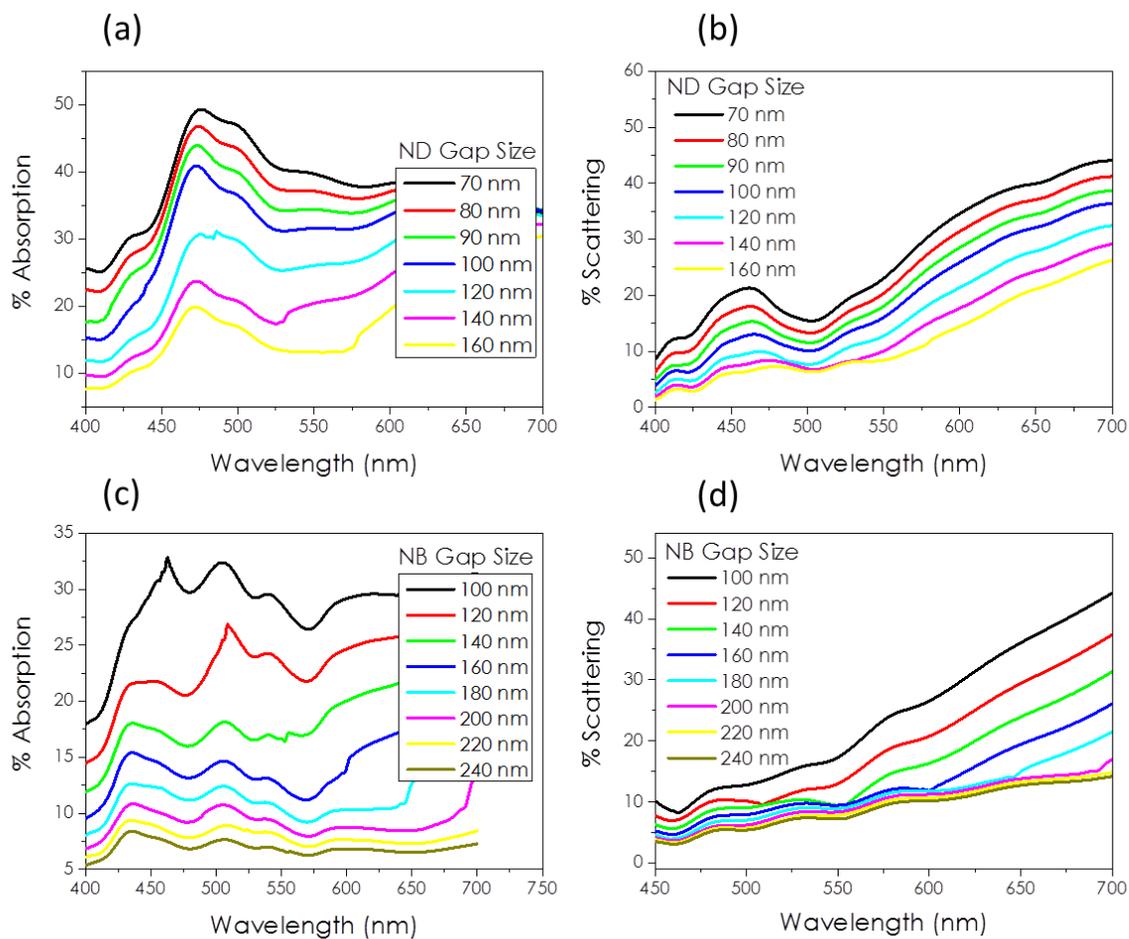

**Figure S3.** (a) Simulated absorption and (b) scattering profiles versus wavelength for the ND arrays with varying gap size from 70 nm to 160 nm. (c) Simulated absorption and (d) scattering profiles versus wavelength for the NB arrays with varying gap size from 100 nm to 250 nm.